\documentclass[reprint,amsmath,amssymb,aps,]{revtex4-2}

\usepackage{graphicx}
\usepackage{dcolumn}
\usepackage{bm}
\usepackage{color} 
\usepackage{amsmath}
\usepackage{enumitem}
\usepackage{amssymb} 
\usepackage{mathrsfs}
\usepackage{hyperref} 
\hypersetup{
	colorlinks=true,
	linkcolor=blue,
	filecolor=blue,      
	citecolor=blue,
	urlcolor=blue  
}

\begin{document}

\title{High-efficiency telecom conversion of heralded atomic biphoton wavepackets}

\author{Ling-Chun Chen, Chang-Wei Lin, Jiun-Shiuan Shiu, Wei-Lin Chen, Yi-Che Wang, and Yong-Fan Chen}

\email{yfchen@mail.ncku.edu.tw}

\affiliation{
$^1$Department of Physics, National Cheng Kung University, Tainan 70101, Taiwan\\ 
$^2$Center for Quantum Frontiers of Research $\&$ Technology, Tainan 70101, Taiwan
}


\date{June 10, 2026}


\begin{abstract}

We demonstrate high-efficiency telecom frequency conversion of heralded atomic biphoton wavepackets using a diamond-type atomic ensemble. By placing a 2.5 MHz heralded-photon spectrum within the high-efficiency region of the converter response, we achieve a conversion efficiency of 79.4(2.6)\% while maintaining strong time-resolved correlations and well-defined temporal wavepackets. For a broader 17.4 MHz input bandwidth, the conversion efficiency is reduced to about 55\%, whereas the temporal waveform remains largely preserved. This behavior reflects the nearly flat central response of the converter, which mainly causes spectral-edge loss rather than temporal-mode distortion. These results identify spectral matching as an effective route to efficient and low-distortion telecom conversion of narrowband quantum light from atomic systems.

\end{abstract}


\maketitle


\newcommand{\FigOne}{
    \begin{figure*}[t]
    \centering
    \includegraphics[width = 16.8 cm]{fig1}
    \caption{
Experimental architecture of the atomic biphoton source and telecom frequency conversion platform. (a), (b) Energy-level schemes of the double-$\Lambda$ SFWM source and diamond-type conversion process. The fields $\Omega_1$, $\Omega_2$, $\hat{a}_p$, $\hat{a}_t$, $\Omega_c$, $\Omega_d$, and $\hat{a}_s$ denote pump 1, pump 2, probe, trigger, coupling, driving, and converted signal, respectively. (c), (d) Corresponding experimental setups. F, filter; DM, dichroic mirror; EFS, etalon filter set; HWP, half-wave plate; L, lens; M, mirror; PBS, polarizing beam splitter; POL, polarizer; QWP, quarter-wave plate; SPCM, single-photon counting module; SMF, single-mode fiber; MMF, multi-mode fiber.
}
    \label{fig:setup}
    \end{figure*}
}

\newcommand{\FigTwo}{
    \begin{figure}[t]
    \centering
    \includegraphics[width = 8.4 cm]{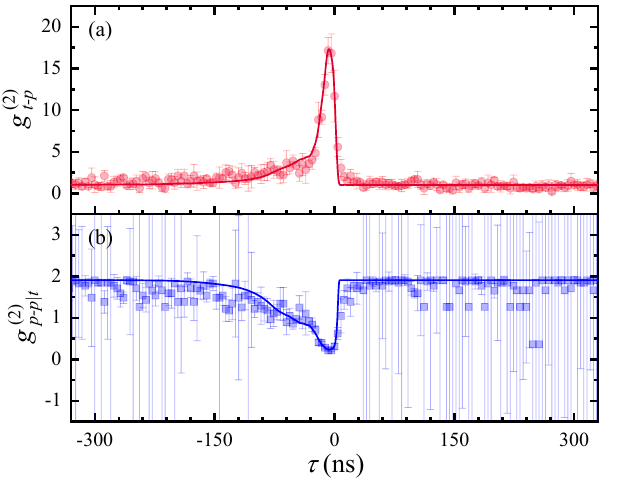}
    \caption{
Benchmarking heralded single photons. (a) Trigger--probe cross-correlation $g^{(2)}_{t\text{-}p}(\tau)$ and (b) conditional autocorrelation $g^{(2)}_{p\text{-}p|t}(\tau)$, showing strong temporal correlations and sub-Poissonian statistics. Symbols and curves denote experiment and theory, respectively. Error bars show one standard deviation. Parameters: ${\rm OD}=8$, $\Omega_1=0.9\Gamma$, $\Omega_2=4\Gamma$, and $\Delta_1=-4\Gamma$.
}
    \label{fig:biphoton}
    \end{figure}
}

\newcommand{\FigThree}{
    \begin{figure}[t]
    \centering
    \includegraphics[width = 8.4 cm]{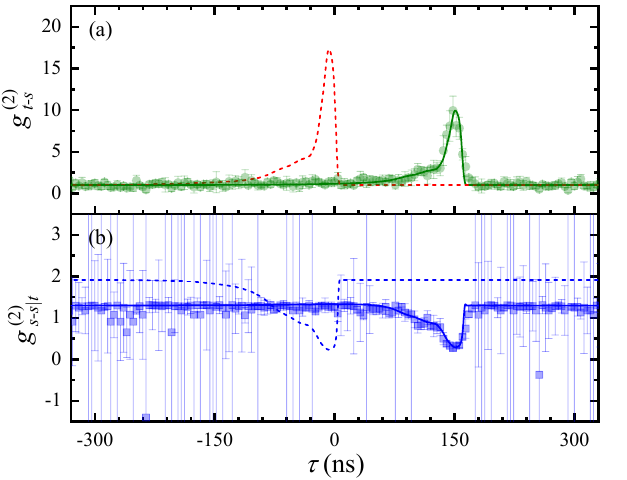}
    \caption{
Telecom conversion of heralded photons. (a) Trigger--signal cross-correlation $g^{(2)}_{t\text{-}s}(\tau)$ and (b) conditional autocorrelation $g^{(2)}_{s\text{-}s|t}(\tau)$, showing preserved temporal correlations and sub-Poissonian statistics. Symbols and curves denote experiment and theory, respectively; dashed curves show the pre-conversion results from Fig.~\ref{fig:biphoton}. Parameters: ${\rm OD}=110$, $\Omega_c=20\Gamma$, $\Omega_d=12\Gamma$, $\Delta_p=-4\Gamma$, $\Delta_c=8\Gamma$, and $\Delta_d=-5\Gamma$.
}
    \label{fig:QFC}
    \end{figure}
}

\newcommand{\FigFour}{
    \begin{figure}[t]
    \centering
    \includegraphics[width = 8.0 cm]{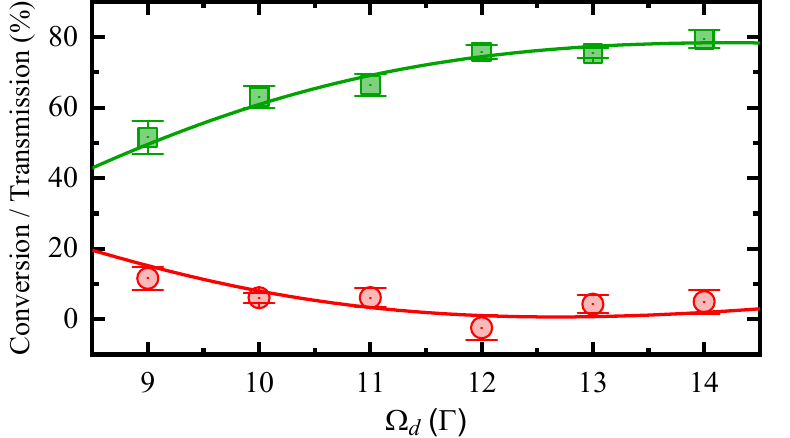}
    \caption{
High-efficiency telecom conversion under spectral matching. Signal conversion efficiency (squares) and probe transmission (circles) versus driving field strength $\Omega_d$. Solid curves show theory. Error bars denote the standard deviation from four independent measurements. Source parameters: ${\rm OD}=17$, $\Omega_1=1.8\Gamma$, $\Omega_2=2\Gamma$, and $\Delta_1=-4\Gamma$, corresponding to a 2.5 MHz input bandwidth. Conversion parameters: ${\rm OD}=120$, $\Omega_c=23\Gamma$, $\Delta_p=-4\Gamma$, $\Delta_c=8\Gamma$, and $\Delta_d=-5\Gamma$.
}
    \label{fig:diamond spectrum}
    \end{figure}
}



%
\textit{Introduction.} Efficient interfaces between narrowband atomic photon sources and low-loss telecom fiber channels are important for long-distance quantum communication, distributed quantum computing, and photonic networking \cite{LDC1,DQC1,LDC2,QKD3,DQC3,LDC3}. Atomic ensembles based on spontaneous four-wave mixing (SFWM) provide a versatile platform for generating narrowband biphotons with long coherence times \cite{SFWM1,SFWM2,SFWM3,Shiu1,SFWM6,50kHz}, suitable for quantum storage and synchronization of remote atomic nodes \cite{QS1,QS2,QS3}. However, most atomic transitions lie in the visible or near-infrared, leading to a wavelength mismatch with low-loss telecom fiber channels.

Several approaches have been explored to generate or access telecom-band quantum light. Atomic cascade configurations and related atom-based biphoton schemes can directly access telecom photons \cite{TBatom1,TBatom2,TBatom3,TBatom5,SB}, but their spectral and temporal properties are largely fixed by atomic structure. Cavity-enhanced nonlinear crystals offer broader tunability \cite{SPDC1,SPDC3}, yet high spectral purity often requires strong filtering or high-finesse cavities, reducing brightness. These considerations motivate an atom-based frequency-conversion approach for narrowband atomic interfaces, with native MHz-bandwidth compatibility despite a larger footprint and limited wavelength flexibility.

Telecom frequency conversion provides such an interface and has been used to connect memory-compatible photons to telecom networks \cite{Maring2018}. In resonant atomic media, near-resonant enhancement can reduce the required pump power and suppress broadband background noise, making this approach especially relevant for atomic quantum nodes and memories \cite{TFC1,TFC2,TFC3,TFC5}. Diamond-type atomic converters are particularly compatible with DLCZ-type architectures, with demonstrated efficiencies above 50\% \cite{CE54,CE65}. Recent experiments have achieved nearly 80\% conversion with weak coherent inputs \cite{CE80}, consistent with theoretical predictions for efficient steady-state operation \cite{PoHan}. Extending this performance to heralded atomic biphotons, however, requires controlling the spectral overlap between the biphoton bandwidth and the finite acceptance window of the converter, because this overlap can affect both the conversion efficiency and the output temporal wavepacket.

In practical systems, the conversion efficiency and output temporal mode depend on the spectral overlap between the input photon and the converter acceptance window \cite{Allgaier2017}. This issue is particularly relevant for heralded SFWM biphotons used in Hong--Ou--Mandel interference \cite{HOM1,HOM2} and Bell-state measurements \cite{BSM2,BSM3}. However, the effect of finite conversion acceptance on the efficiency and temporal correlations of heralded atomic biphotons has not been experimentally clarified.

In this work, we demonstrate high-efficiency telecom conversion of heralded atomic biphoton wavepackets using a diamond-type atomic ensemble. By placing a 2.5 MHz heralded-photon spectrum within the high-efficiency region of the converter response, we achieve a conversion efficiency of 79.4(2.6)\%. For a broader 17.4 MHz input bandwidth, the efficiency is reduced to about 55\%, while the converted photons still retain strong temporal correlations and well-defined wavepackets. This comparison shows that incomplete spectral overlap mainly reduces the converted photon number, rather than causing significant distortion of the dominant temporal mode. These results highlight spectral matching as a practical route to efficient and low-distortion telecom conversion of narrowband quantum light.


\FigOne



%
\textit{Experimental setup.} Heralded single photons are generated via double-$\Lambda$ SFWM in a cold $^{87}$Rb ensemble, as shown in Fig.~\ref{fig:setup}(a). The process is driven by two pump fields: a far-detuned 795 nm laser with Rabi frequency $\Omega_1$ and detuning $\Delta_1$, and a resonant 780 nm laser with Rabi frequency $\Omega_2$. These pumps address the $\sigma^-$ and $\sigma^+$ transitions and generate a $\sigma^-$-polarized probe photon $\hat{a}_p$ together with a $\sigma^+$-polarized trigger photon $\hat{a}_t$.

The probe photon $\hat{a}_p$ is converted to the telecom band via a diamond-type configuration [Fig.~\ref{fig:setup}(b)]. Unlike the multi-Zeeman generation stage, the conversion process operates on an isolated transition with atoms optically pumped into $|5S_{1/2}, F=2, m_F=-2\rangle$. This configuration stabilizes the population distribution and suppresses unwanted optical pumping, improving the reproducibility of the conversion process. We apply a $\sigma^-$-polarized driving field ($\Omega_d$, 1324 nm) and a $\sigma^-$-polarized coupling field ($\Omega_c$, 780 nm) to realize this single-transition scheme. The $\sigma^+$-polarized probe field carries the SFWM detuning $\Delta_p = \Delta_1$. Spin-angular-momentum conservation then yields a $\sigma^+$-polarized telecom signal field $\hat{a}_s$ at 1367 nm. The conversion efficiency is optimized by tuning the one-photon detunings $(\Delta_d,\Delta_c)$ and the two-photon detuning $\delta = \Delta_p+\Delta_d$.

The biphoton source is implemented in a cold-atom ensemble [Fig.~\ref{fig:setup}(c)], while the telecom converter is realized in a spatially separated ensemble [Fig.~\ref{fig:setup}(d)]. Either correlated photon can in principle be converted. However, converting the trigger photon requires tuning $\Omega_2$, which changes the nonlinear coupling strength and spectral response, leading to waveform distortion and reduced pairing ratios \cite{Shiu2}. In contrast, tuning $\Omega_1$ mainly shifts the probe spectrum while preserving stable SFWM operation and strong correlations. This flexibility allows the input bandwidth to be tailored for frequency conversion without noticeably degrading the source performance. We therefore choose the probe photon as the conversion input.

The probe is sent through a 20 m fiber to the distant converter, adding a $\sim$100 ns delay for strictly heralded operation. Within the biphoton temporal waveform, the trigger photon is emitted after the probe photon. The delay therefore allows the trigger photon to be detected before the probe photon enters the converter. The detection event projects the probe field onto a heralded single-photon state with clear antibunching \cite{Shiu3}, providing the heralded input for the conversion measurement.

To operate in the genuine single-photon regime, we upgraded the filtering and collection paths. In the visible path, a polarizing beam splitter and bandpass filters provide 136 dB isolation, with a 24\% collection efficiency before detection by a Si-based single-photon counting module (SPCM-AQRH-13-FC: dark counts $\sim 50$ s$^{-1}$, quantum efficiency $\sim 65\%$, dead time 20 ns, jitter 350 ps). In the telecom path, spectral and spatial filtering suppress the driving field by 135 dB. The converted photons are coupled into a single-mode fiber with a 57\% efficiency and detected by an InGaAs SPCM (PDM-IR: dark counts $\sim 1000$ s$^{-1}$, quantum efficiency $\sim 21\%$, dead time 200~$\mu$s, jitter 130 ps). With a 4 ns coincidence window, the overall post-conversion optical efficiency equals the internal conversion efficiency multiplied by this 57\% collection efficiency. Temporal correlations between the trigger and converted signal photons are recorded using a time-of-flight multiscaler. The experiment operates at 100 Hz with a 10 ms cycle. During the measurement window, the cooling and magnetic fields are switched off. Biphoton generation is driven by a 5~$\mu$s pulse of $\Omega_1$ after optical pumping with $\Omega_2$, while $\Omega_c$ and $\Omega_d$ cover the probe arrival, with $\Omega_c$ applied 3~$\mu$s earlier.


%
\textit{Theoretical model.} To benchmark the measured quantum correlations, we employ a microscopic open-quantum-system model for the coupled atomic and optical fields~\cite{Shiu1,PoHan}. Solving the coupled equations yields the trigger and probe field operators, $\hat{a}_t$ and $\hat{a}_p$. We define the photon generation rate as $R_l=(c/L)\langle\hat{a}_l^\dagger\hat{a}_l\rangle$ ($l=t,p$), and the normalized trigger--probe cross-correlation function as
\begin{align}
	g_{t\text{-}p}^{(2)}(\tau)
	=& 1+
	\frac{(L/c)^2}{R_{t}R_{p}}
	\left|
	\langle\hat{a}_t^\dagger(t)\hat{a}_p^\dagger(t+\tau)\rangle
	\right|^2.
	\label{eq1}
\end{align}
In Eq.~(\ref{eq1}), the unity term accounts for accidental coincidences, while the second term represents the biphoton temporal wavepacket associated with the correlated emission process. A dominant contribution from this term indicates strong quantum correlations and a clear departure from classical statistics.

The cross-correlation $g_{t\text{-}p}^{(2)}(\tau)$ determines the purity of heralded single photons. A key metric is the zero-delay conditional autocorrelation $g_{p\text{-}p|t}^{(2)}(\tau)$ \cite{Shiu3}, which in the ideal limit is
\begin{align}
	g_{p\text{-}p|t}^{(2)}(\tau)
	=
	\left[4g_{t\text{-}p}^{(2)}(\tau)-2\right]
	\left[g_{t\text{-}p}^{(2)}(\tau)\right]^{-2}.
	\label{eq2}
\end{align}
A larger $g_{t\text{-}p}^{(2)}$ lowers $g_{p\text{-}p|t}^{(2)}$ toward zero, indicating sub-Poissonian statistics. At long delays, temporal correlation vanishes and $g_{t\text{-}p}^{(2)}$ approaches unity; consequently, $g_{p\text{-}p|t}^{(2)}$ approaches 2, reflecting thermal statistics of uncorrelated fields.

To account for experimental imperfections, including leakage and detector dark counts, we introduce the channel purity $P_l$ for each detection channel. This quantity is defined as the ratio between genuine SFWM signal counts and the total detected counts. The measured cross-correlation function is then modified as
\begin{align}
	g_{t\text{-}p}^{(2)}(\tau)
	\rightarrow
	P_{t}P_{p}\left[g_{t\text{-}p}^{(2)}(\tau)-1\right]+1.
	\label{eq3}
\end{align}
The conditional autocorrelation must be corrected accordingly. Expressed in terms of the ideal cross-correlation, the measured value becomes
\begin{align}
	g_{p\text{-}p|t}^{(2)}(\tau)
	\rightarrow
	\frac{1+P_{p}^2+2P_{t}P_{p}\left(1+P_{p}\right)\left[g_{t\text{-}p}^{(2)}(\tau)-1\right]}
	{\left\{P_{t}P_{p}\left[g_{t\text{-}p}^{(2)}(\tau)-1\right]+1\right\}^2}.
	\label{eq4}
\end{align}
Detailed derivations are provided in the Supplemental Material.

\FigTwo



%
\textit{Heralded photons from atomic biphotons.} Figure~\ref{fig:biphoton}(a) shows the normalized trigger--probe cross-correlation $g^{(2)}_{t\text{-}p}(\tau)$ for an optical depth (OD) of 8, $\Omega_1=0.9\Gamma$, $\Omega_2=4\Gamma$, $\Delta_1=-4\Gamma$, and $\gamma_{21}=0.001\Gamma$, where $\Gamma = 2\pi \times 6$~MHz. The theoretical prediction agrees well with experiment. Error bars represent the standard deviation from four independent 15 min measurements. After correcting for optical leakage and detector dark counts, the peak cross-correlation reaches 18, confirming strong nonclassical temporal correlations.

\FigThree

Under these conditions, the calculated probe and trigger generation rates are $R_p = 8.9 \times 10^5$~s$^{-1}$ and $R_t = 1.4 \times 10^6$~s$^{-1}$, with a pair rate of $7.3 \times 10^5$~s$^{-1}$. The corresponding pairing ratios are 0.82 and 0.52, respectively. This asymmetry arises from unequal Clebsch--Gordan coefficients, yielding different probe- and trigger-photon rates for the same correlated-pair rate.

To verify the single-photon character of the heralded probe photons, we evaluate the conditional autocorrelation $g^{(2)}_{p\text{-}p|t}(\tau)$, as shown in Fig.~\ref{fig:biphoton}(b). The minimum value reaches 0.22, confirming sub-Poissonian photon statistics. Fluctuations outside the correlation window originate from low accidental coincidence counts and have negligible influence on the conversion measurements. The well-defined temporal wavepacket and narrow bandwidth of the heralded probe photons therefore provide a controlled input for examining how the finite spectral acceptance of the converter affects telecom conversion.


%
\textit{Telecom conversion with spectral matching.} Probe photons at 795 nm are converted to the 1367 nm telecom band via diamond-type four-wave mixing. At ${\rm OD}=110$, $\Omega_c=20\Gamma$, $\Omega_d=12\Gamma$, and $(\Delta_p,\Delta_c,\Delta_d)=(-4,8,-5)\Gamma$, we observe a peak cross-correlation of $g^{(2)}_{t\text{-}s} \approx 10$ [Fig.~\ref{fig:QFC}(a)]. Although this value is reduced from the pre-conversion value of $\sim$18 mainly because of the higher dark-count rate of the InGaAs detector and residual optical leakage, it still shows strong temporal correlations after conversion. Using channel purities $P_t=0.89$ ($\sim$10.2\% leakage, $\sim$0.3\% dark counts) and $P_s=0.54$ ($\sim$21.9\% leakage, $\sim$23.8\% dark counts), Eq.~(\ref{eq3}) predicts a peak value of 9, in reasonable agreement with experiment. This agreement indicates that the reduced correlation peak is mainly due to a reduced signal-to-background ratio caused by increased background counts, rather than by significant distortion of the quantum wavepacket. The conditional autocorrelation of the converted telecom photons reaches a minimum of 0.27 [Fig.~\ref{fig:QFC}(b)], remaining well below the classical limit.

The converted wavepacket has a full width at half maximum of $\sim$20 ns, corresponding to a Fourier-limited bandwidth of $\sim$17.4 MHz, consistent with the pre-conversion result. The signal conversion efficiency, defined here as the internal photon-to-photon conversion probability, is $\sim$55\%, below the nearly 80\% efficiency obtained with weak coherent inputs. This reduction arises from incomplete spectral overlap between the probe photons and the finite conversion acceptance, with a bandwidth of $\sim$40 MHz \cite{CE80}. Because the converter response is nearly flat near its high-efficiency center, the dominant spectral components are converted with similar relative weights, while edge components mainly reduce the converted photon number. The temporal waveform therefore remains largely unchanged, indicating that finite spectral acceptance mainly causes photon loss rather than temporal-mode distortion.

\FigFour

To improve the conversion efficiency, we reduce the SFWM bandwidth to $\sim$2.5~MHz by tuning the source parameters to ${\rm OD}=17$, $\Omega_1=1.8\Gamma$, and $\Omega_2=2\Gamma$, while maintaining a photon generation rate on the order of $10^7$~s$^{-1}$. With the probe spectrum lying within the high-efficiency region of the conversion response, and with the conversion ensemble optimized to ${\rm OD}=120$ and $\Omega_c=23\Gamma$, the signal conversion efficiency reaches 79.4(2.6)\% at $\Omega_d=14\Gamma$ [Fig.~\ref{fig:diamond spectrum}]. This result shows that spectral overlap with the high-efficiency region of the converter response is crucial for approaching the steady-state conversion efficiency. The observed efficiency is close to the steady-state value expected for the present OD, and previous theory indicates that increasing the OD can further improve the efficiency toward the ideal limit~\cite{PoHan}.

The preservation of the temporal waveform is important for interference-based quantum-networking protocols. A well-defined wavepacket supports temporal-mode matching in Hong--Ou--Mandel interference, while the narrow bandwidth provides a long coherence length that relaxes path-length stability requirements in fiber systems. Together with the achieved conversion efficiency and spectral controllability, these properties make the platform suitable for narrowband telecom interfaces based on atomic quantum light.



%
\textit{Conclusion.} We demonstrate high-efficiency telecom conversion of heralded atomic biphoton wavepackets in a diamond-type atomic ensemble, reaching a signal conversion efficiency of 79.4(2.6)\%. The converted photons retain strong temporal correlations, sub-Poissonian statistics, and well-defined wavepackets. Comparing 2.5 MHz and 17.4 MHz inputs shows that finite spectral acceptance mainly reduces the converted photon number, while the nearly flat central converter response suppresses significant distortion of the dominant temporal mode. These results establish an atomic interface for low-distortion telecom conversion of narrowband quantum light.



\textit{Acknowledgements.} We thank Y.-T. Ma, M.-Y. Lin, and X.-Q. Zhong for contributions to the initial experimental setup. This work was supported by the National Science and Technology Council of Taiwan under Grants No. 114-2112-M-006-007 and No. 114-2119-M-007-012. Support from the Center for Quantum Science and Technology under the Higher Education Sprout Project of the Ministry of Education in Taiwan is also acknowledged.



\onecolumngrid 
\newpage

\setcounter{equation}{0}
\setcounter{figure}{0}
\setcounter{table}{0}
\setcounter{page}{1}
\makeatletter
\renewcommand{\theequation}{S\arabic{equation}}
\renewcommand{\thefigure}{S\arabic{figure}}
\renewcommand{\thetable}{S\arabic{table}}
\renewcommand{\bibnumfmt}[1]{[S#1]}
\renewcommand{\citenumfont}[1]{S#1}

\makeatother

\begin{center}
    \textbf{\large Supplemental Material for\\ High-efficiency telecom conversion of heralded atomic biphoton wavepackets}
    \vspace{0.3cm}
    
    Ling-Chun Chen, Chang-Wei Lin, Jiun-Shiuan Shiu, Wei-Lin Chen, Yi-Che Wang, and Yong-Fan Chen$^*$
    \vspace{0.2cm}
    
    \textit{\small $^1$Department of Physics, National Cheng Kung University, Tainan 70101, Taiwan\\ 
    $^2$Center for Quantum Frontiers of Research $\&$ Technology, Tainan 70101, Taiwan}
    \vspace{0.5cm}
\end{center}

\begin{center}
    \begin{minipage}{0.78\textwidth} 
        \small 
        \quad
        This Supplemental Material presents the derivation of the corrected conditional autocorrelation function $g_{p\text{-}p|t}^{(2)}(\tau)$ under realistic experimental conditions. To account for background counts and detector dark noise, the detected fields are modeled as a linear superposition of statistically independent signal and noise operators. By introducing channel purities ($P_t, P_p$) that quantify the fraction of genuine biphoton events, the coincidence expectation values are systematically evaluated. The resulting analytical expression relates the measured correlation function to the intrinsic quantum statistics and correctly recovers the noise-free limit.
    \end{minipage}
\end{center}
\vspace{2em} 


\maketitle

To quantify the single-photon purity of the heralded probe field, we evaluate the conditional autocorrelation function defined as~\cite{ShiuS1}
\begin{align}
g_{p\text{-}p|t}^{(2)}(\tau)
=
\frac{
\langle\hat{a}_t^\dagger(t)\hat{a}_t(t)\rangle
\langle\hat{a}_t^\dagger(t)\hat{a}_{p1}^\dagger(t+\tau)
       \hat{a}_{p2}^\dagger(t+\tau)\hat{a}_{p2}(t+\tau)
       \hat{a}_{p1}(t+\tau)\hat{a}_t(t)\rangle
}{
\langle\hat{a}_t^\dagger(t)\hat{a}_{p1}^\dagger(t+\tau)\hat{a}_{p1}(t+\tau)\hat{a}_t(t)\rangle
\langle\hat{a}_t^\dagger(t)\hat{a}_{p2}^\dagger(t+\tau)\hat{a}_{p2}(t+\tau)\hat{a}_t(t)\rangle}.
\label{S1}
\end{align}
Here $p1$ and $p2$ denote the two probe detection channels in the Hanbury Brown--Twiss (HBT) configuration. The delay $\tau$ represents the relative detection time between the probe and trigger photons. In the ideal limit where only correlated photon pairs are detected, the conditional autocorrelation reduces to $g_{p\text{-}p|t}^{(2)}(\tau)=[4g_{t\text{-}p}^{(2)}(\tau)-2][g_{t\text{-}p}^{(2)}(\tau)]^{-2}$. The second-order cross-correlation function is $g_{t\text{-}p}^{(2)}(\tau)=\langle\hat{a}_t^\dagger\hat{a}_p^\dagger\hat{a}_p\hat{a}_t\rangle\langle\hat{a}_t^\dagger\hat{a}_t\rangle^{-1}\langle\hat{a}_p^\dagger\hat{a}_p\rangle^{-1}$.

To account for experimental imperfections in the correlation measurements, we decompose the annihilation operator of each detection channel $l \in \{t, p1, p2\}$ into signal and noise contributions, $\hat{a}_l = \hat{a}_{l,b} + \hat{a}_{l,e}$. Here the subscript $b$ denotes the genuine biphoton signal, while $e$ represents environmental noise such as background light and detector dark counts. The biphoton signal and environmental noise are assumed to be statistically independent, and noise contributions in different spatial channels are also independent. Consequently, the cross-term expectation values vanish, $\langle\hat{a}_{l,b}^\dagger\hat{a}_{l,e}\rangle =\langle\hat{a}_{l,e}^\dagger\hat{a}_{l,b}\rangle =\langle\hat{a}_{l,b}\hat{a}_{l,e}\rangle =\langle\hat{a}_{l,b}^\dagger\hat{a}_{l,e}^\dagger\rangle = 0$, and $\langle\hat{a}_{m,e}^\dagger\hat{a}_{n,e}\rangle=0$ ($m\neq n$).

With the statistical independence between the biphoton signal and environmental noise established, we now evaluate the expectation values appearing in the conditional autocorrelation function of Eq.~\eqref{S1}. Substituting the decomposed operators for each detection channel and removing the vanishing cross terms yields the following expressions:
\begin{itemize}[label=$\blacksquare$, leftmargin=*]
\item \textbf{Trigger-channel intensity} $\langle\hat{a}_t^\dagger\hat{a}_t\rangle$
\begin{align}
\langle\hat{a}_t^\dagger\hat{a}_t\rangle
=
\langle(\hat{a}_{t,b}^\dagger+\hat{a}_{t,e}^\dagger)(\hat{a}_{t,b}+\hat{a}_{t,e})\rangle
=
\langle\hat{a}_{t,b}^\dagger\hat{a}_{t,b}\rangle
+\langle\hat{a}_{t,e}^\dagger\hat{a}_{t,e}\rangle.
\label{S2}     
\end{align}

\item \textbf{Two-fold coincidence (probe channel 1)} $\langle\hat{a}_t^\dagger\hat{a}_{p1}^\dagger\hat{a}_{p1}\hat{a}_t\rangle$
\begin{align}
&\langle\hat{a}_t^\dagger\hat{a}_{p1}^\dagger\hat{a}_{p1}\hat{a}_t\rangle
=
\langle(\hat{a}_{t,b}^\dagger+\hat{a}_{t,e}^\dagger)(\hat{a}_{p1,b}^\dagger+\hat{a}_{p1,e}^\dagger)(\hat{a}_{p1,b}+\hat{a}_{p1,e})(\hat{a}_{t,b}+\hat{a}_{t,e})\rangle \nonumber\\
&=
\langle(\hat{a}_{t,b}^\dagger\hat{a}_{p1,b}^\dagger+\hat{a}_{t,b}^\dagger\hat{a}_{p1,e}^\dagger+\hat{a}_{t,e}^\dagger\hat{a}_{p1,b}^\dagger+\hat{a}_{t,e}^\dagger\hat{a}_{p1,e}^\dagger)
    (\hat{a}_{p1,b}\hat{a}_{t,b}+\hat{a}_{p1,b}\hat{a}_{t,e}+\hat{a}_{p1,e}\hat{a}_{t,b}+\hat{a}_{p1,e}\hat{a}_{t,e})\rangle
\nonumber\\
&=
\langle\hat{a}_{t,b}^\dagger\hat{a}_{p1,b}^\dagger\hat{a}_{p1,b}\hat{a}_{t,b}\rangle+\langle\hat{a}_{t,b}^\dagger\hat{a}_{p1,e}^\dagger\hat{a}_{p1,e}\hat{a}_{t,b}\rangle 
    +\langle\hat{a}_{t,e}^\dagger\hat{a}_{p1,b}^\dagger\hat{a}_{p1,b}\hat{a}_{t,e}\rangle+\langle\hat{a}_{t,e}^\dagger\hat{a}_{p1,e}^\dagger\hat{a}_{p1,e}\hat{a}_{t,e}\rangle
\nonumber\\
&=
\langle\hat{a}_{t,b}^\dagger\hat{a}_{p1,b}^\dagger\hat{a}_{p1,b}\hat{a}_{t,b}\rangle
+
\langle\hat{a}_{t,b}^\dagger\hat{a}_{t,b}\rangle
\langle\hat{a}_{p1,e}^\dagger\hat{a}_{p1,e}\rangle
+
\langle\hat{a}_{t,e}^\dagger\hat{a}_{t,e}\rangle
\langle\hat{a}_{p1,b}^\dagger\hat{a}_{p1,b}\rangle
+
\langle\hat{a}_{t,e}^\dagger\hat{a}_{t,e}\rangle
\langle\hat{a}_{p1,e}^\dagger\hat{a}_{p1,e}\rangle
\nonumber\\
&=
\langle\hat{a}_{t,b}^\dagger\hat{a}_{t,b}\rangle
\langle\hat{a}_{p1,b}^\dagger\hat{a}_{p1,b}\rangle g_{t\text{-}p}^{(2)}
+
\langle\hat{a}_{t,b}^\dagger\hat{a}_{t,b}\rangle
\langle\hat{a}_{p1,e}^\dagger\hat{a}_{p1,e}\rangle
+
\langle\hat{a}_{t,e}^\dagger\hat{a}_{t,e}\rangle
\langle\hat{a}_{p1,b}^\dagger\hat{a}_{p1,b}\rangle
+
\langle\hat{a}_{t,e}^\dagger\hat{a}_{t,e}\rangle
\langle\hat{a}_{p1,e}^\dagger\hat{a}_{p1,e}\rangle
\nonumber\\
&=
(\langle\hat{a}_{t,b}^\dagger\hat{a}_{t,b}\rangle
 +\langle\hat{a}_{t,e}^\dagger\hat{a}_{t,e}\rangle)
(\langle\hat{a}_{p1,b}^\dagger\hat{a}_{p1,b}\rangle
 +\langle\hat{a}_{p1,e}^\dagger\hat{a}_{p1,e}\rangle)
+
\langle\hat{a}_{t,b}^\dagger\hat{a}_{t,b}\rangle
\langle\hat{a}_{p1,b}^\dagger\hat{a}_{p1,b}\rangle
[g_{t\text{-}p}^{(2)}-1]
.
\label{S3}  
\end{align}
To simplify the notation, we introduce the mean photon numbers $n_{l,x}=\langle\hat{a}_{l,x}^\dagger\hat{a}_{l,x}\rangle$, where $l\in\{t,p1,p2\}$ denotes the detection channel and $x\in\{b,e\}$ distinguishes the biphoton signal from environmental noise. We also define the channel purity $P_l=n_{l,b}/(n_{l,b}+n_{l,e})$, which quantifies the fraction of genuine biphoton events in the detected counts. With these definitions, the two-fold coincidence rate in Eq.~\eqref{S3} can be written as $\langle\hat{a}_t^\dagger\hat{a}_{p1}^\dagger\hat{a}_{p1}\hat{a}_t\rangle=(n_{t,b}+n_{t,e})(n_{p1,b}+n_{p1,e})\{P_tP_{p1}[g_{t\text{-}p}^{(2)}-1]+1\}$.

\item \textbf{Two-fold coincidence (probe channel 2)} $\langle\hat{a}_t^\dagger\hat{a}_{p2}^\dagger\hat{a}_{p2}\hat{a}_t\rangle$
\begin{align}
&\langle\hat{a}_t^\dagger\hat{a}_{p2}^\dagger\hat{a}_{p2}\hat{a}_t\rangle
=
\langle(\hat{a}_{t,b}^\dagger+\hat{a}_{t,e}^\dagger)(\hat{a}_{p2,b}^\dagger+\hat{a}_{p2,e}^\dagger)(\hat{a}_{p2,b}+\hat{a}_{p2,e})(\hat{a}_{t,b}+\hat{a}_{t,e})\rangle
\nonumber\\
&=
(n_{t,b}+n_{t,e})(n_{p2,b}+n_{p2,e}) \{ P_t P_{p2} [ g_{t\text{-}p}^{(2)} - 1 ] + 1 \}
.
\label{S4}  
\end{align}
In a standard HBT configuration, the 50/50 beam splitter ensures that the two probe detection paths have identical statistical properties and noise levels. The corresponding channel purities are therefore equal. For notational simplicity we denote $n_{p1,b}=n_{p2,b}\equiv n_{p,b}$, $n_{p1,e}=n_{p2,e}\equiv n_{p,e}$, and $P_{p1}=P_{p2}\equiv P_p$.

\item \textbf{Three-fold coincidence} $\langle\hat{a}_t^\dagger\hat{a}_{p1}^\dagger\hat{a}_{p2}^\dagger\hat{a}_{p2}\hat{a}_{p1}\hat{a}_t\rangle$
\begin{align}
&\langle\hat{a}_t^\dagger\hat{a}_{p1}^\dagger\hat{a}_{p2}^\dagger\hat{a}_{p2}\hat{a}_{p1}\hat{a}_t\rangle
=
\langle(\hat{a}_{t,b}^\dagger+\hat{a}_{t,e}^\dagger)(\hat{a}_{p1,b}^\dagger+\hat{a}_{p1,e}^\dagger)(\hat{a}_{p2,b}^\dagger+\hat{a}_{p2,e}^\dagger)(\hat{a}_{p2,b}+\hat{a}_{p2,e})(\hat{a}_{p1,b}+\hat{a}_{p1,e})(\hat{a}_{t,b}+\hat{a}_{t,e})\rangle
\nonumber\\
&=
\langle\hat{a}_{t,b}^\dagger\hat{a}_{p1,b}^\dagger\hat{a}_{p2,b}^\dagger\hat{a}_{p2,b}\hat{a}_{p1,b}\hat{a}_{t,b}\rangle 
\nonumber\\
&\quad
+\langle\hat{a}_{t,b}^\dagger\hat{a}_{p1,b}^\dagger\hat{a}_{p1,b}\hat{a}_{t,b}\rangle
 \langle\hat{a}_{p2,e}^\dagger\hat{a}_{p2,e}\rangle
+\langle\hat{a}_{t,b}^\dagger\hat{a}_{p2,b}^\dagger\hat{a}_{p2,b}\hat{a}_{t,b}\rangle
 \langle\hat{a}_{p1,e}^\dagger\hat{a}_{p1,e}\rangle
+\langle\hat{a}_{p1,b}^\dagger\hat{a}_{p2,b}^\dagger\hat{a}_{p2,b}\hat{a}_{p1,b}\rangle
 \langle\hat{a}_{t,e}^\dagger\hat{a}_{t,e}\rangle
\nonumber\\
&\quad
+\langle\hat{a}_{t,b}^\dagger\hat{a}_{t,b}\rangle
 \langle\hat{a}_{p1,e}^\dagger\hat{a}_{p1,e}\rangle
 \langle\hat{a}_{p2,e}^\dagger\hat{a}_{p2,e}\rangle 
+\langle\hat{a}_{p1,b}^\dagger\hat{a}_{p1,b}\rangle
 \langle\hat{a}_{t,e}^\dagger\hat{a}_{t,e}\rangle
 \langle\hat{a}_{p2,e}^\dagger\hat{a}_{p2,e}\rangle 
+\langle\hat{a}_{p2,b}^\dagger\hat{a}_{p2,b}\rangle
 \langle\hat{a}_{t,e}^\dagger\hat{a}_{t,e}\rangle
 \langle\hat{a}_{p1,e}^\dagger\hat{a}_{p1,e}\rangle
\nonumber\\
&\quad
+\langle\hat{a}_{t,e}^\dagger\hat{a}_{t,e}\rangle\langle\hat{a}_{p1,e}^\dagger\hat{a}_{p1,e}\rangle\langle\hat{a}_{p2,e}^\dagger\hat{a}_{p2,e}\rangle
\nonumber\\
&=
[4g_{t\text{-}p}^{(2)}-2][g_{t\text{-}p}^{(2)}]^{-2}
\langle\hat{a}_{t,b}^\dagger\hat{a}_{p1,b}^\dagger\hat{a}_{p1,b}\hat{a}_{t,b}\rangle
\langle\hat{a}_{t,b}^\dagger\hat{a}_{p2,b}^\dagger\hat{a}_{p2,b}\hat{a}_{t,b}\rangle
\langle\hat{a}_{t,b}^\dagger\hat{a}_{t,b}\rangle^{-1}
\nonumber\\
&\quad+
[2n_{t,b}n_{p,b}n_{p,e}g_{t\text{-}p}^{(2)}+2n_{p,b}^2n_{t,e}]
+
(n_{t,b}n_{p,e}^2+2n_{p,b}n_{t,e}n_{p,e})+n_{t,e}n_{p,e}^2
\nonumber\\
&=
4n_{t,b}n_{p,b}^2[g_{t\text{-}p}^{(2)}-1]+2n_{t,b}n_{p,b}^2
\nonumber\\
&\quad+
2n_{t,b}n_{p,b}n_{p,e}[g_{t\text{-}p}^{(2)}-1]
+
2n_{t,b}n_{p,b}n_{p,e}
+
2n_{p,b}^2n_{t,e}
+
n_{t,b}n_{p,e}^2+2n_{p,b}n_{t,e}n_{p,e}+n_{t,e}n_{p,e}^2
\nonumber\\
&=
2n_{t,b}n_{p,b}(2n_{p,b}+n_{p,e})[g_{t\text{-}p}^{(2)}-1]
+
(n_{t,b}+n_{t,e})[n_{p,b}^2+(n_{p,b}+n_{p,e})^2]
.
\label{S5}
\end{align}
\end{itemize}
Substituting Eqs.~\eqref{S2}--\eqref{S5} into Eq.~\eqref{S1} gives the zero-delay conditional autocorrelation function
\begin{align}
g_{p\text{-}p|t}^{(2)}=&
\frac{
(n_{t,b}+n_{t,e})\{2n_{t,b}n_{p,b}(2n_{p,b}+n_{p,e})[g_{t\text{-}p}^{(2)}-1]+(n_{t,b}+n_{t,e})[n_{p,b}^2+(n_{p,b}+n_{p,e})^2]\}
}{
(n_{t,b}+n_{t,e})^2(n_{p,b}+n_{p,e})^2 \{ P_t P_{p} [ g_{t\text{-}p}^{(2)} - 1 ] + 1 \}^2
}
\nonumber\\
=&
\frac{1+P_p^2+2P_tP_p(1+P_p)\left[g_{t\text{-}p}^{(2)}-1\right]}
{\left\lbrace P_t P_{p} \left[g_{t\text{-}p}^{(2)}-1\right] + 1 \right\rbrace^2},
\label{S6}
\end{align}
which is consistent with Eq.~\eqref{eq4} of the main text.



\begin{thebibliography}{99}
	

\newcommand{\enquote}[1]{``#1''}

\bibitem{LDC1}
L.-M. Duan, M. D. Lukin, J. I. Cirac, \emph{et al.},
\enquote{Long-distance quantum communication with atomic ensembles and linear optics,}
Nature \textbf{414}, 413 (2001).

\bibitem{DQC1} 
H.-S. Zhong, H. Wang, Y.-H. Deng, \emph{et al.}, 
\enquote{Quantum computational advantage using photons,} 
Science \textbf{370}, 1460 (2020).

\bibitem{LDC2}
X.-M. Hu, C.-X. Huang, Y.-B. Sheng, \emph{et al.}, 
\enquote{Long-distance entanglement purification for quantum communication,} 
Phys. Rev. Lett. \textbf{126}, 010503 (2021).

\bibitem{QKD3} 
Y. Liu, W.-J. Zhang, C. Jiang, \emph{et al.},
\enquote{Experimental twin-field quantum key distribution over 1000 km fiber distance,}
Phys. Rev. Lett. \textbf{130}, 210801 (2023).

\bibitem{DQC3} 
D. Main, P. Drmota, D. P. Nadlinger, \emph{et al.},
\enquote{Distributed quantum computing across an optical network link,} 
Nature \textbf{638}, 383 (2025).

\bibitem{LDC3} 
M. Pittaluga, Y. S. Lo, A. Brzosko, \emph{et al.}, 
\enquote{Long-distance coherent quantum communications in deployed telecom networks,} 
Nature \textbf{640}, 911 (2025).

\bibitem{SFWM1}
V. Balić, D. A. Braje, P. Kolchin, \emph{et al.}, 
\enquote{Generation of paired photons with controllable waveforms,} 
Phys. Rev. Lett. \textbf{94}, 183601 (2005).

\bibitem{SFWM2}
L. Zhao, X. Guo, C. Liu, \emph{et al.}, 
\enquote{Photon pairs with coherence time exceeding 1 $\mu$s,} 
Optica \textbf{1}, 84 (2014).

\bibitem{SFWM3}
C.-Y. Hsu, Y.-S. Wang, J.-M. Chen, \emph{et al.}, 
\enquote{Generation of sub-MHz and spectrally-bright biphotons from hot atomic vapors with a phase mismatch-free scheme,} 
Opt. Express \textbf{29}, 4632 (2021).

\bibitem{Shiu1}
J.-S. Shiu, Z.-Y. Liu, C.-Y. Cheng, \emph{et al.},
\enquote{Observation of highly correlated ultrabright biphotons through increased atomic ensemble density in spontaneous four-wave mixing,}
Phys. Rev. Res. \textbf{6}, L032001 (2024).

\bibitem{SFWM6} 
J.-K. Lin, T.-H. Chien, C.-T. Wu, \emph{et al.}, 
\enquote{Observation of subnatural-linewidth biphotons in a two-level atomic ensemble,} 
Phys. Rev. Lett. \textbf{134}, 043602 (2025).

\bibitem{50kHz} 
Y.-S. Wang, K.-B. Li, C.-F. Chang, \emph{et al.},
\enquote{Temporally ultralong biphotons with a linewidth of 50 kHz,}
APL Photonics \textbf{7}, 126102 (2022).

\bibitem{QS1}
K. S. Choi, H. Deng, J. Laurat, \emph{et al.}, 
\enquote{Mapping photonic entanglement into and out of a quantum memory,} 
Nature \textbf{452}, 67 (2008).

\bibitem{QS2} 
D. S. Ding, Z. Y. Zhou, B. S. Shi, \emph{et al.}, 
\enquote{Single-photon-level quantum image memory based on cold atomic ensembles,} 
Nat. Commun. \textbf{4}, 2527 (2013).

\bibitem{QS3} 
Y.-W. Cho, G. T. Campbell, J. L. Everett, \emph{et al.}, 
\enquote{Highly efficient optical quantum memory with long coherence time in cold atoms,}
Optica \textbf{3}, 100 (2016).

\bibitem{TBatom1} 
T. Chanelière, D. N. Matsukevich, S. D. Jenkins, \emph{et al.},
\enquote{Quantum telecommunication based on atomic cascade transitions,}
Phys. Rev. Lett. \textbf{96}, 093604 (2006).

\bibitem{TBatom2} 
R. T. Willis, F. E. Becerra, L. A. Orozco, \emph{et al.}, 
\enquote{Photon statistics and polarization correlations at telecommunications wavelengths from a warm atomic ensemble,}
Opt. Express \textbf{19}, 14632 (2011).

\bibitem{TBatom3} 
W. Zhang, D.-S. Ding, S. Shi, \emph{et al.}, 
\enquote{Storing a single photon as a spin wave entangled with a flying photon in the telecommunication bandwidth,} 
Phys. Rev. A \textbf{93}, 022316 (2016).

\bibitem{TBatom5} 
P.-Y. Tu, C.-Y. Hsu, W.-K. Huang, \emph{et al.},
\enquote{Temporally long C-band heralded single photons generated from hot atoms,}
APL Photon. \textbf{10}, 106104 (2025).

\bibitem{SB}
Z.-Y. Liu, J.-S. Shiu, W.-L. Chen, \emph{et al.},
\enquote{Microscopic origin of superradiant biphoton emission in atomic ensembles,}
arXiv:2602.11438 (2026).

\bibitem{SPDC1}
K. Niizeki, K. Ikeda, M. Zheng, \emph{et al.},
\enquote{Ultrabright narrow-band telecom two-photon source for long-distance quantum communication,}
Appl. Phys. Express \textbf{11}, 042801 (2018).

\bibitem{SPDC3}
M.-Y. Gao, Y.-H. Li, Y. Li, \emph{et al.},
\enquote{Narrowband telecom-band polarization-entangled photon source by superposed monolithic cavities,}
Phys. Rev. A \textbf{109}, 033720 (2024).

\bibitem{Maring2018} 
N. Maring, D. Lago-Rivera, A. Lenhard, \emph{et al.},
\enquote{Quantum frequency conversion of memory-compatible single photons from 606 nm to the telecom C-band,}
Optica \textbf{5}, 507 (2018).

\bibitem{TFC1} 
R. T. Willis, F. E. Becerra, L. A. Orozco, \emph{et al.}, 
\enquote{Four-wave mixing in the diamond configuration in an atomic vapor,} 
Phys. Rev. A \textbf{79}, 033814 (2009).

\bibitem{TFC2} 
H. Jeong, H. Kim, J. Bae, \emph{et al.},
\enquote{Doppler-broadened four-wave mixing under double-resonance optical pumping in the $5S_{1/2}$--$5P_{3/2}$--$4D_{5/2}$ transition of warm $^{87}$Rb atoms,}
Opt. Express \textbf{29}, 42384 (2021).

\bibitem{TFC3} 
W.-H. Zhang, Y.-H. Ye, L. Zeng, \emph{et al.}, 
\enquote{Telecom-wavelength conversion in a high optical depth cold atomic system,} 
Opt. Express \textbf{31}, 8042 (2023).

\bibitem{TFC5} 
J. A. Rowland, C. Perrella, R. F. Offer, \emph{et al.}, 
\enquote{Characterization of near-infrared to telecom frequency conversion in a rubidium-filled hollow-core photonic-crystal fiber,} 
Opt. Express \textbf{33}, 18076 (2025).

\bibitem{CE54} 
A. G. Radnaev, Y. O. Dudin, R. Zhao, \emph{et al.},
\enquote{A quantum memory with telecom-wavelength conversion,}
Nature Phys. \textbf{6}, 894 (2010).

\bibitem{CE65}
Y. O. Dudin, A. G. Radnaev, R. Zhao, \emph{et al.},
\enquote{Entanglement of light-shift compensated atomic spin waves with telecom light,}
Phys. Rev. Lett. \textbf{105}, 260502 (2010).

\bibitem{CE80}
L.-C. Chen, M.-Y. Lin, J.-S. Shiu, \emph{et al.},
\enquote{High-efficiency telecom frequency conversion via a diamond-type atomic ensemble,}
Phys. Rev. A \textbf{112}, 013709 (2025).

\bibitem{PoHan} 
P.-H. Tseng, L.-C. Chen, J.-S. Shiu, \emph{et al.}, 
\enquote{Quantum interface for telecom frequency conversion based on diamond-type atomic ensembles,} 
Phys. Rev. A \textbf{109}, 043716 (2024).

\bibitem{Allgaier2017} 
M. Allgaier, V. Ansari, L. Sansoni, \emph{et al.}, 
\enquote{Highly efficient frequency conversion with bandwidth compression of quantum light,} 
Nat. Commun. \textbf{8}, 14288 (2017).

\bibitem{HOM1}
C. K. Hong, Z. Y. Ou, and L. Mandel, 
\enquote{Measurement of subpicosecond time intervals between two photons by interference,} 
Phys. Rev. Lett. \textbf{59}, 2044 (1987).

\bibitem{HOM2}
H. Ollivier, S. E. Thomas, S. C. Wein, \emph{et al.}, 
\enquote{Hong-Ou-Mandel interference with imperfect single photon sources,} 
Phys. Rev. Lett. \textbf{126}, 063602 (2021).

\bibitem{BSM2}
E. Arenskötter, S. Kucera, O. Elshehy, \emph{et al.}, 
\enquote{Full Bell-basis measurement of an atom-photon 2-qubit state and its application for quantum networks,} 
Phys. Rev. Res. \textbf{6}, 023061 (2024).

\bibitem{BSM3}
N. Hauser, M. J. Bayerbach, S. E. D’Aurelio, \emph{et al.}, 
\enquote{Boosted Bell-state measurements for photonic quantum computation,} 
npj Quantum Inf. \textbf{11}, 41 (2025).

\bibitem{Shiu2}
J.-S. Shiu, C.-W. Lin, Y.-C. Huang, \emph{et al.},
\enquote{Frequency-tunable biphoton generation via spontaneous four-wave mixing,} 
Phys. Rev. A \textbf{110}, 063723 (2024).

\bibitem{Shiu3}
J.-S. Shiu, C.-W. Lin, and Y.-F. Chen, 
\enquote{Asymmetric biphoton generation under ground-state decoherence and phase mismatch in a cold atomic ensemble,}
Adv. Quantum Technol. \textbf{8}, e2500052 (2025).


\end{thebibliography}

\begin{thebibliography}{99}
	
\bibitem{ShiuS1}
J.-S. Shiu, C.-W. Lin, and Y.-F. Chen,
Asymmetric biphoton generation under ground-state decoherence and phase mismatch in a cold atomic ensemble,
Adv. Quantum Technol. \textbf{8}, e2500052 (2025).

\end{thebibliography}
\end{document}